\documentclass[useAMS,usenatbib]{mn2e}

\title[Jupiter and Super-Earth embedded in a gaseous disc]
{Jupiter and Super-Earth embedded in a gaseous disc}
\author[E. Podlewska and E. Szuszkiewicz]{E. Podlewska$$\thanks{E-mail:
edytap@univ.szczecin.pl (EP)} and E. Szuszkiewicz$$\thanks{E-mail: 
szusz@univ.szczecin.pl (ES)} \\
Institute of Physics and CASA*, University of Szczecin, ul. Wielkopolska 15,
70-451 Szczecin, Poland}
\begin{document}

\date{}

\pagerange{\pageref{firstpage}--\pageref{lastpage}} \pubyear{}

\maketitle

\label{firstpage}

\begin{abstract}
In this paper we investigate the evolution of a pair of interacting 
planets - a Jupiter mass planet and a Super-Earth with the mass 5.5 
$M_{\oplus}$ - orbiting a Solar type star and embedded in a gaseous 
protoplanetary disc.  We focus on the effects of type I and  II orbital  
migrations, caused by the planet-disc interaction, leading to the 
Super-Earth capture in first order mean motion resonances by the Jupiter. 
The stability of the resulting resonant system in which the Super-Earth 
is on the internal orbit relatively to the Jupiter has been studied 
numerically by means of full 2D hydrodynamical simulations. Our main 
motivation is to determine the Super-Earth behaviour in the presence of 
the gas giant in the system. It has been found that the Jupiter captures 
the Super-Earth into the interior 3:2 or 4:3 mean motion resonances and 
the stability of such configurations depends on the initial planet positions 
and eccentricity evolution. If the initial separation of planet orbits is 
larger or close to  that required for the exact resonance than the final 
outcome is the migration of the pair of planets with the rate similar to 
that of the gas giant at least for time of our simulations.  Otherwise we 
observe a scattering of the Super-Earth from the disc.  The evolution of 
planets immersed in the gaseous disc has been compared with their behaviour 
in the case of the classical three-body problem when the disc is absent.
\end{abstract}

\section{Introduction}
More than ten years have passed since the discovery of first extrasolar 
low mass planets (with masses 0.02, 4.3 and 3.9 $M_{\oplus}$) around 
pulsar PSR B1257+12 \citep{wolfra}
and the first hot Jupiter orbiting 51 Pegasi, a Solar-type star 
\citep{mayque}. 
Since then,  more than two hundred gas giants have been found around 
main sequence stars and only four low 
(less  massive than 10  $M_{\oplus}$) mass planets.
The first among these four, with a mass 7.5 $M_{\oplus}$ orbiting the 
nearby star
Gliese 876 has been discovered by 
the high-precision radial velocity monitoring at the W. M.
Keck Observatory in Hawaii \citep{rivera}. The second one, 
OGLE-2005-BLG-390Lb 
with a mass of 5.5 $M_{\oplus}$, 
has been found by using  the gravitational
microlensing technique  
by OGLE, PLANETS and MOA collaborations \citep{beaulieu}.  
Finally the last two, with  masses about 5 and 8 $M_{\oplus}$,
has been observed around star Gliese 581 \citep{udry} thanks to HARPS 
(High Accuracy Radial Velocity for Planetary Searcher), a very precise
spectrograph located on the ESO telescope at La Silla in Chile. 
All three stars 
 harboring low mass planets are M dwarfs.
The planets
in this mass range are often called Super-Earths and we will adopt
this name here. These few examples
are just a preview of what is to come 
soon. The microlensing and the high-precision radial-velocity 
planet-search programmes 
can bring new discoveries at any time. 
Even a network of small ground-based
telescopes is able to increase the  number of known Earth-like planets
using the Transit Timing Variation (TTV) method.
The French-European mission COROT, which is 
capable to detect low mass planets, is already operating and
collecting data. In 2009 the Kepler 
satellite will be launched 
and also will search for Earth-size planets. 
In other words,
it is a perfect time to have a closer look at the  
possible planet configurations formed as a result of 
the evolution of young planetary systems in which different mass 
planets might coexist with each other.  
Here we focus on the resonant configurations between two planets - 
a gas giant (we will call it Jupiter) and a Super-Earth,
orbiting a Solar-type star. 
Our prediction of the occurrence of the commensurabilities in such systems
may turn out to be particularly valuable in the case of the TTV observations,
because the differences in the time intervals between successive transits, 
caused by planet-planet
interactions, are largest near mean-motion resonance 
\citep{agol}. 

It has been recognized that resonant structures
may form as a result of large scale orbital migration in young planetary
systems due to tidal interactions between planets or protoplanets and a 
gaseous disc, in which the whole system is still embedded.
The different mass objects, which we expect to find in forming
planetary systems, will migrate with different rates. The final 
configurations will
depend on the intricate interplay among many physical processes including 
planet-planet, disc-planet and planet-star interactions. 
One scenario is that  
the convergent migration brings the
giant planets closer to each other and they can become locked in low
order commensurability \citep{bryden, kley} 
as it is observed at least in three multiplanet
systems (GJ 876, HD 82943 and 55 Cnc).  
Also, the low mass planets can 
undergo the convergent
migration and form a resonant structure \citep{papszusz}. The pulsar
planets around PSR B1257+12 might be an outcome of such scenario. 
The differences in the migration rates of low and high mass 
planets may  also lead to the convergent
migration \citep{hawa}. We 
will  consider this type of convergent migration here as a mechanism 
for capture of a Super-Earth by a Jupiter into  mean motion
commensurabilities and we will follow the pair of planets further in
their evolution to establish the stability of their configuration.
It has been found already by 
\citet{thommes}
that for typical protoplanetary disc parameters the low mass
planets can be captured by massive planet into exterior  resonances.
The resonant captures play also an important role in modelling terrestrial
planet formation and their survival in the presence of migrating
gas giant.
The early divergent conclusions on the occurrence of terrestrial planets 
in hot Jupiter systems  \citep{levison, armitage, 
raymond05, mansig} have been
clarified by most recent studies 
\citep{fonel05, 
fonel07a, fonel07b, zhou, raymond, mandell} which predict that 
terrestrial planets can grow and be retained in the hot-Jupiter systems
both interior and exterior of the gas giant.
The most relevant feature of these investigations for the present study
is the possibility of the
existence of terrestrial planets on the internal orbits relative to Jupiter. 
Such  planets could become
captured in the mean motion commensurability which is then maintained
during their
evolution so an  outcome in this case will be the Jupiter
and the low-mass planet migrating together towards the star.

Our aim here is to consider in full detail
the evolution of the close pair of planets:
a gas giant on the external orbit and a Super-Earth on the internal one
in the young planetary system, when both planets are embedded in a 
gaseous disc.
 We have performed 2D
hydrodynamical  simulations in order to 
determine the occurrence of the first order
mean motion resonances in such systems and at the same time to examine 
the possible planet configurations as the outcome of the convergent
orbital migration. 

Our paper is organized as follows.
In Section 2 we summarize shortly few facts 
about two types of migration,
which we will
use in the present study. 
In Section 3 we  discuss the Jupiter and Super-Earth
system in the context of
the restricted three body problem when the disc is absent. Moreover we
have performed
simple N-body calculations in order to determine the Super-Earth
dynamics in a presence of Jupiter and compare it with the classical
results of celestial mechanics.
This provides us with a well studied and understood framework in which
we can present  the results of our hydrodynamical simulations of a 
young planetary system, where planets are still embedded in a gaseous disc.
In Section 4 we 
describe  our hydrodynamical simulations and show the 
planetary orbit evolution together with the changes in the protoplanetary
disc structure. 
The  discussion and our conclusions are given
in Section 5.

\section{Slow and fast migration}
\label {section2}
Migration due to planet-disc interaction might play an important 
role in shaping up planet configurations in the planetary systems
as we have already mentioned in the previous section. 
From the dynamical point of view one of the most important consequences of 
this process is an
occurrence of planets 
locked in mean-motion resonances and this will be our main concern here. 

\begin{figure}
\vskip 8cm
\includegraphics{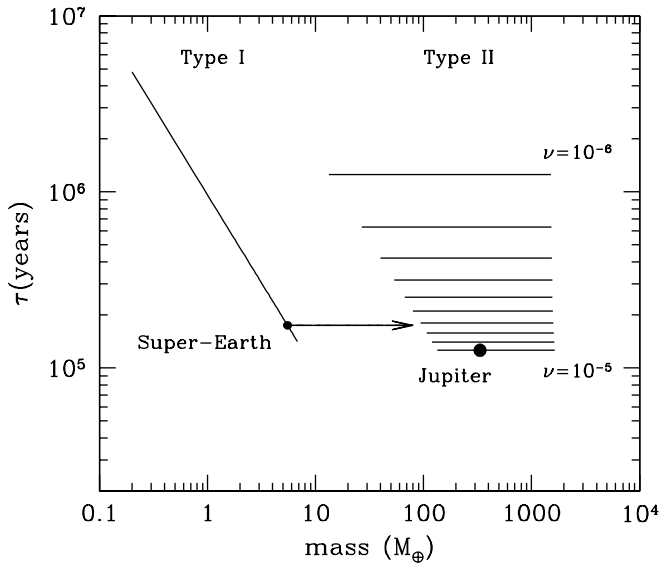}
\caption{\label{fig1}{
The comparison between 
the times of migration for planets with different masses  at 5.2 AU. 
We have concentrated 
on classical type I and II migration cases, for which the theoretical 
and numerical evaluations agreed very well and their understanding is 
rather established \citep{ward97, paplar, papter06, bate03, dangelo03}. 
}}
\end{figure}

The migration rates for different planet masses has been estimated 
by number of authors, see the review by \cite{fivebig}. 
Their results have been illustrated in Fig.~\ref{fig1} where we plot the
migration time of a planet as a function of its mass. There are two
mass regimes which are of interest here, namely (0.1 - 30$M_{\oplus}$)
and (150 - 1500$M_{\oplus}$) for which in a typical protoplanetary disc
we can talk about two different types of migration, called simply type I
and type II respectively. 

The migration time for low mass planets embedded
in a gaseous disc (type I migration) has been derived by \cite{tanaka02} 
in the form
\begin{eqnarray}
\tau_{I}=(2.7+1.1 \gamma)^{-1} \frac{M_*}{m_p}\frac{M_*}{\Sigma_p
{r_p}^2} 
\left( H \over r \right)^2
{\Omega_p}^{-1}    
\label{tanaka}
\end{eqnarray} 
Here $m_p$ is mass of the planet,
$r_p$ is the distance from the central star $M_*$, $\Sigma_p$ is the disc
surface density,  $H/r$ and 
$\Omega_p$ are  the disc aspect ratio  and angular
velocity respectively. The coefficient $\gamma$ depends on the disc surface
density profile, which is expressed as $\Sigma(r) \propto r^{-\gamma}$. 
Assuming $\gamma=0$ for the flat surface density
distribution, $\Sigma_p= 2000$ kg/m$^2$, $ H/r =0.05$ 
 and $r_p=5.2 AU$ we have calculated
the migration time as a function of the planet mass and draw it in
Fig.~\ref{fig1}.  

Type II migrators open a gap in the disc and their evolution 
is determined by 
the radial velocity drift in the disc
\begin{equation}
v_r = {3\over 2}{\nu \over r_p}
\end{equation}
where $\nu$ is a kinematic viscosity parameter.
The migration time can be estimated as \citep{linpap93} 
 
\begin{eqnarray}
\tau_{II}={r_p\over v_r} =\frac{2 {r_p}^2}{3 \nu}
\label{tauii}
\end{eqnarray}
It has been shown in Fig.~\ref{fig1} 
for $r_p=5.2 AU$ and different values of $\nu$  
($10^{-5},
2\cdot 10^{-5}, 3\cdot 10^{-5}, ..., 9\cdot 10^{-5}$ and $10^{-6}$)
expressed in the
dimensionless units discussed in Section~\ref{section41}.
The migration time in this formulation
does not depend on the planet mass. In drawing lines 
for a given viscosity parameter
we have taken into account the condition for a gap opening in the disc which 
reads
\begin{equation}
{m_p \over M_*} > {40\nu \over r_p^2 \Omega_p}
\label{gap}
\end{equation}
and from which it is clear that the bigger is viscosity  $\nu$ the bigger
is the mass of the planet, which is able to open a gap.
Recently, \cite{edgar} has argued that similarly like in the case of type
I migration also here the migration time  depends  
on the planet
mass and the disc surface density. His relation gives for our set of 
parameters somewhat faster migration
for the Jupiter in comparison with the equation (\ref{tauii}), so it does 
not change our argument here. 

We have chosen the mass of the type I and type II migrators to be 
5.5 $M_{\oplus}$ and 1 Jupiter mass respectively. Their locations  have been
marked in Fig.~\ref{fig1} accordingly to their masses  and in the case of 
the Jupiter also to the viscosity adopted in the disc ($\nu  = 10^{-5}$).
It is obvious from this figure that we should expect a 
convergent migration
if the Jovian type planet is at the external orbit, further 
away from the star, and the Super-Earth at the internal one. We 
have checked that for the reverse configuration we are getting the 
divergent migration. 
For the disc with the low viscosity this situation may change and the 
Jupiter mass planet would migrate slower than the Super-Earth. In order
to get the convergent migration the Super-Earth should than be located 
at the exterior orbit exactly as the Earth considered in \citet{thommes}. 
The overall picture is quite simple and straightforward. For a given disc
profile and masses of the planets, the relative speed of type I and type 
II migrators is controlled
by the viscosity in the disc. However, the details of this scenario require
more attention. We will come back to this issue in Section~\ref{section5}.

Starting from  the configuration illustrated in Fig.~\ref{fig1}
we have investigated the possible 
resonances in the system containing a star and two planets with masses, 
which are much different from each other. Their mass ratio to the central
star is equal to 
$m_{SE}= 1.65 \cdot 10^{-5}$ for the Super-Earth
and $m_J = 1 \cdot 10^{-3}$ for the Jupiter, so their masses differ from 
each
other by almost two orders of magnitude and  
the Super-Earth, in the absence of the disc,
could be treated at least approximately
in the context of the classical restricted three-body
problem, where one body is so light that it does not influence the
motion of the two others. In the next section we compare the dynamics of 
our system with the classical results of the three-body problem. 

\section{Classical three-body problem}
\label{section3}

Before tackling the full problem of the Jupiter and Super-Earth, orbiting
around Solar-type star, embedded in
a gaseous disc let us first consider a simpler well studied situation in
celestial mechanics when the disc is not present in the system. We are 
interested in motion
of three celestial bodies under their mutual gravitational attraction.  
In our system the mass of two planets is much smaller than the central 
star mass and the mass of the Super-Earth is significantly smaller than
that of the Jupiter.

The dynamics of a pair of interacting planets orbiting a star has been 
studied in full details by \citet{gladman}.
The main conclusion of his paper 
is that for initially circular planetary orbits with the semi-major axes
$a$ and $a(1+\Delta)$, the relative orbital separation $\Delta$
between the two planets with the mass ratios to the mass of the central
star equal $m_1$ and $m_2$, respectively,  has to  satisfy the 
following criterion
\begin{equation}
\Delta > 2.4 (m_1 + m_2)^{1/3}
\end{equation}
in order to be Hill stable. This means that for the planetary orbit separations
fulfilling this condition planets will not suffer from the close encounters
at any time. The close encounter will result in the scattering of the 
Super-Earth, which may leave the system, fall onto the star or the Jupiter. 
In our case, for the Jupiter ($m_1 =m_J$) and
Super-Earth ($m_2 =m_{SE}$) considered in this paper this criterion gives 
\begin{equation}
\Delta > 0.24,
\label{glad024}
\end{equation}
which is the same as for the Jupiter and a massless body, since 
$m_{SE}/m_J \ll 1$.
We have found how good is this simple analytical prediction in our case 
by performing direct orbit integration.
Employing a simple N-body code,  
which uses Bulirsch-Stoer integrator and proved itself to
be a convenient way to obtain high-accuracy solutions of the differential
equations (see e.g. \citet{terqupap}), 
we have calculated
the planetary orbits in the region where  the interior 3:2, 4:3 and 5:4 Jupiter 
mean-motion resonances are located. We have put the Super-Earth at the
fixed location $a_{SE}=1$, defining in this way our unit of distance, and the 
Jupiter initial semi-major axes have been varied in the
range from 1.15 till 1.45 in these units.
The initial eccentricities of both planets have
been set to zero. Our results 
have been summarized in Table~\ref{tab1}, where  the number
of conjunctions for each initial configuration
that the system survives without suffering a close encounter is given in the
column 3.  When the Jupiter is located at the distance 1.24 or smaller 
we have observed the orbit crossing and scattering of the Super-Earth.
Our result agrees very well with the stability criterion given by 
\cite{gladman}.  

\begin{figure}
\vskip 6.5cm
\includegraphics{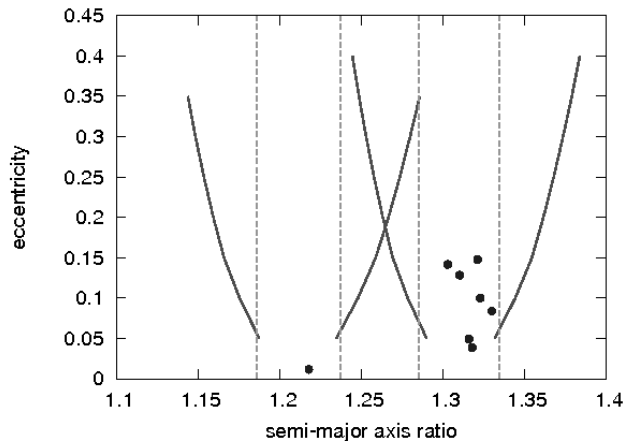}
\caption{\label{fig2}{Maximum libration regions in (semi-major axis ratio,
eccentricity) space for the 3:2 (right) and 4:3 (left) resonances. The 
semi-major axis ratio has been calculated dividing the semi-major axis
of Jupiter by the semi-major axis of the small body. The eccentricity is
that of the small body. The dashed line is the resonance width for 
zero eccentricity and the solid line is the libration width derived using
the pendulum approach. The dots correspond to the locations  
of asteroids in the 3:2 (Hilda group) and 4:3
(Thule group) mean motion resonances with Jupiter in the Solar System. 
We plot Thule asteroid and Hilda, Babylon,
Normannia, Xanthomalitia, 2000VV39, 2001PR1 and 2002DL3 from Hilda group
using data from
the Near-Earth Object Website hosted by The Jet Propulsion Laboratory.
}}
\end{figure}

Hill stability criterion is useful for predicting close encounters between
planets but is not telling us much about other properties of motion as
for example  whether or not the system is chaotic.
\cite{wisdom80} basing on the resonance overlapping criterion found 
a boundary
between chaotic and stable motions in the form
\begin{equation}
\Delta  > 1.5 a_J m^{2/7}
\label{wisdom}
\end{equation}
where $m$ denotes mass ratio of the two biggest bodies in the system.
The coefficient in front of the mass in the original paper was 2, but
it has been improved later by numerical simulations \citep{duncan} and we 
have adopted here the more recent value 1.5.
The occurrence of the mean motion resonances is therefore
very important for stability of the system. 
The width of the resonance regions for the circular
restricted three-body problem has been derived by \cite{wisdom80} (see
also \cite{lecar}). In Fig.~\ref{fig2} we have plotted the width of 3:2 
and 4:3 interior resonances for zero eccentricity given by  \cite{lecar} 
for the Jupiter (vertical
dashed lines). In this Figure the semi-major axis ratio is obtained 
by dividing the semi-major axis of the Jupiter by that of the small
body.
\cite{winmur} have discussed the various analytical models that have
been used in the study of interior first order resonances.
Analyzing simple pendulum model they derived formula describing maximum
deviation of semi-major axis from the nominal value given by exact mean
motion commensurability. Their results for 3:2 and 4:3 resonances are
presented also in Fig.~\ref{fig2} for comparison (solid lines). 
The theoretical prediction
illustrated by the pendulum model (similar results were obtained using 
Hamiltonian approach) that for higher value of the eccentricity
of the small body the first order resonance width is larger, 
has not been verified by direct integration 
\citep{winmur}. The size of the resonant libration
region derived by \cite{winmur} performing numerical integrations of the full
equations of motion is quite similar to that given by \cite{lecar}. 
The plot of maximum libration widths has been supplemented with a few
examples of the mean motion commensurability known 
in our Solar System.
The dots denote Hilda and Thule groups of
asteroids. This groups are particularly interesting
because they are locked in the mean motion resonances with Jupiter. Note
that asteroids are located exactly within regions predicted by analysis of the
restricted three-body problem.

We have performed an analysis of the chaotic regions for our system (the
Super-Earth and Jupiter around the Solar-type star) in the
similar way it has been done for asteroid belt. First, let us evaluate the size
of the chaotic region using equation (\ref{wisdom}).
Substituting in this equation for $m$ the mass ratio of  Jupiter and Sun we
have found that the small body (here the Super-Earth) should be outside the
region of large scale chaos if 
\begin{equation}
\Delta  >  0.21a_J.
\end{equation}
\begin{figure}
\vskip 21.5cm
\includegraphics{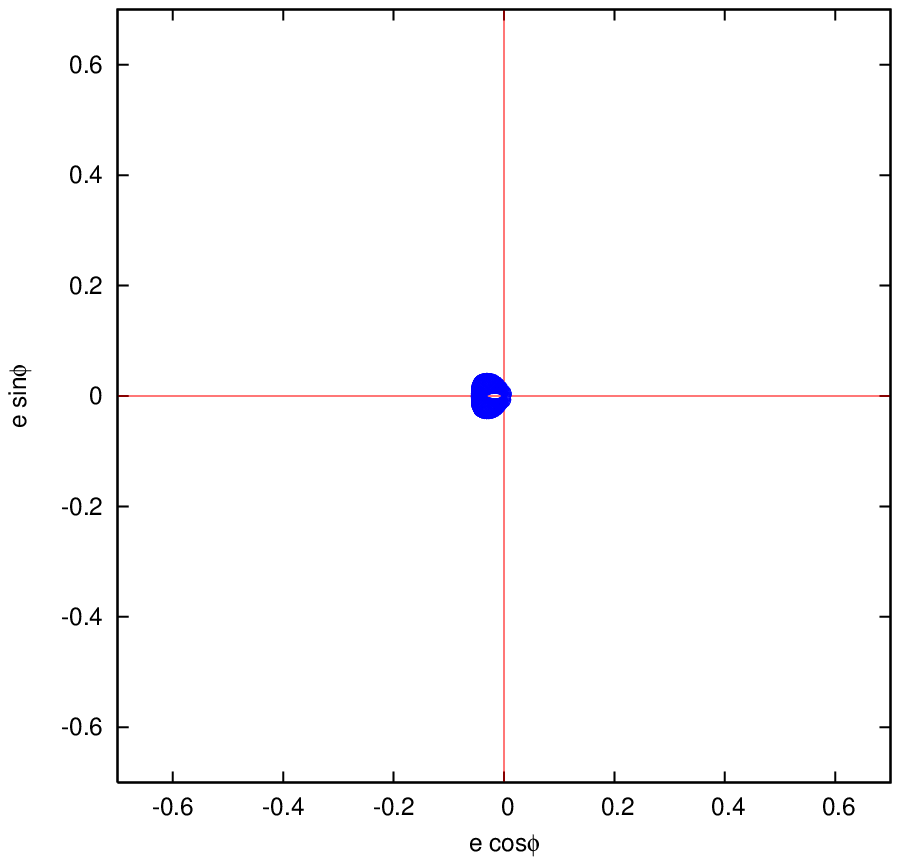}
\includegraphics{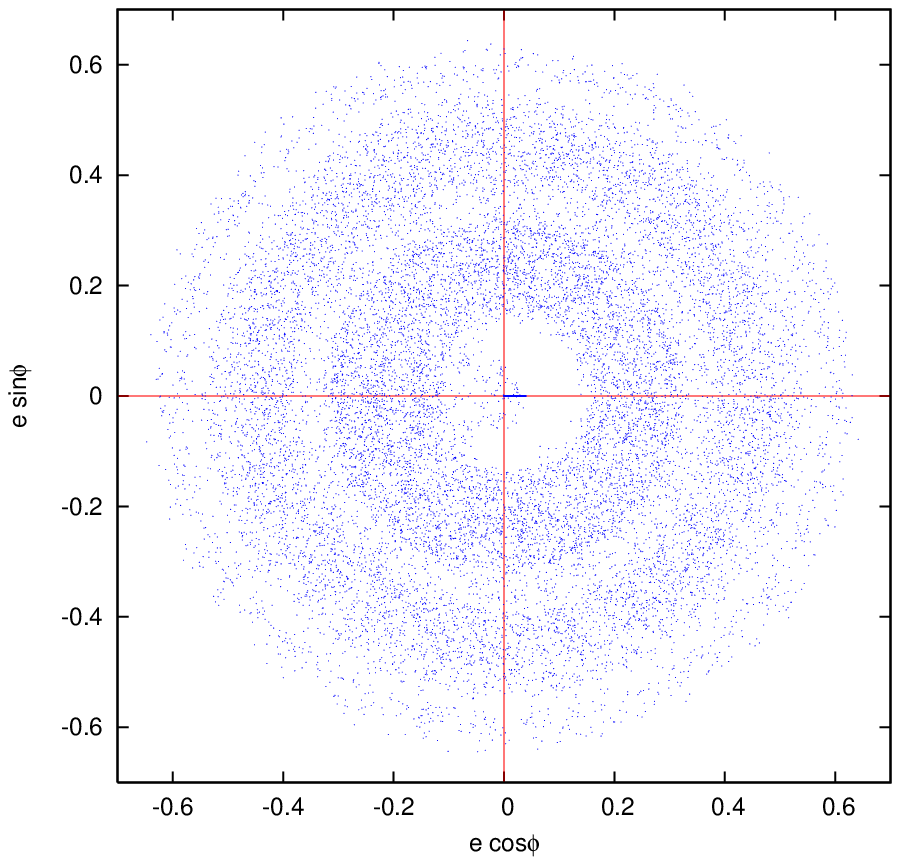}
\includegraphics{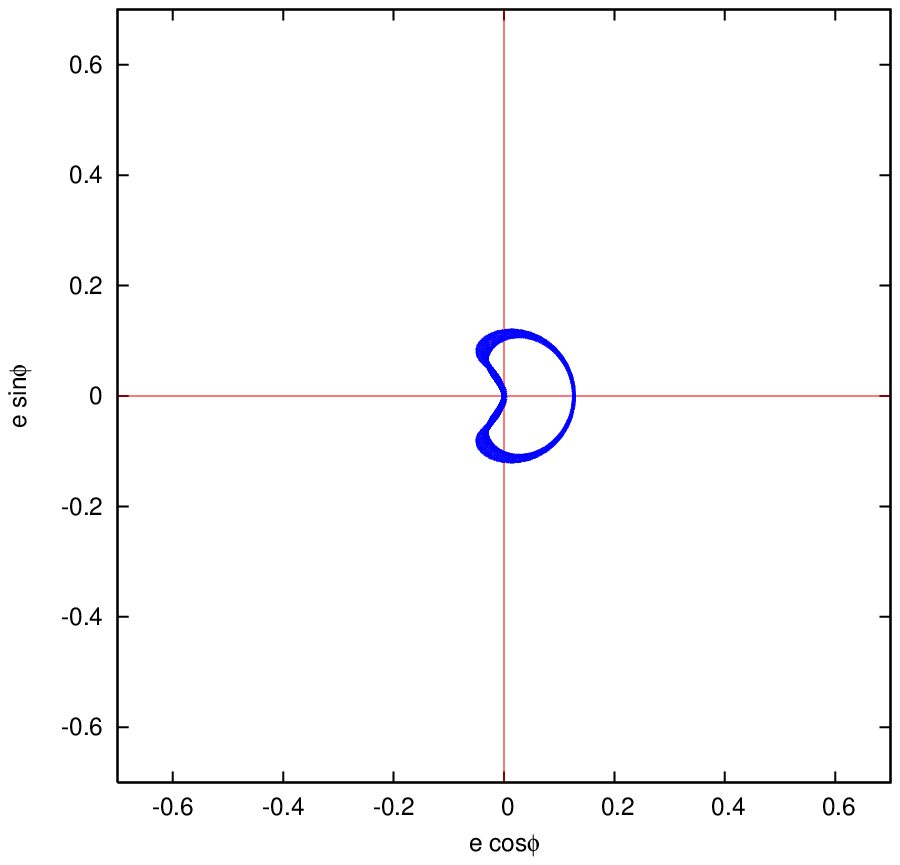}
\caption{\label{fig3}{The structure of phase space trajectories of 
the Super-Earth in three different planet configuration. From the top  
to the bottom panels, the Jupiter initial semi-major axis is 
1.27, 1.28 and 1.30 correspondingly. The distance from the origin is
measured by the Super-Earth eccentricity, $e$ and the polar angle is given
by the resonance phase $\phi$. 
}}
\end{figure}
According to this criterion the Super-Earth would be in the region of 
large
scale chaos when the Jupiter's semi-major axis is smaller than 1.27. 

Next, we have examined the properties of the planetary orbits in the 
Hill stable region, which means for the Jupiter semi-major axis having value 
in the range
between 1.24 and 1.45. 
We have found two  families of regular orbits separated from
each other by the zone of the chaotic behaviour, which is located between
1.277 and 1.289. 
It should be pointed out  here that
the fact of the existence of the chaotic zone outside the region 
indicated
by the resonance overlap criterion is not entirely unexpected. It should
be kept in mind that 
criterion given by equation (\ref{wisdom}) is quite approximate and 
Wisdom's 
scaling 
law is strictly 
valid
only in the asymptotic limit, when the integer $p$ defining first order
resonance $(p+1) : p$ is much bigger than one \citep{malhotra}. In our
case of the Jupiter mass planet we are concerned with the first order
resonances with small $p$.

The families of the stable orbits can be nicely portrayed drawing the
structure of the phase space trajectories in a plane on which the polar
coordinates are ($e$, $\phi$) where $e$ is the Super-Earth
eccentricity and $\phi$ is the resonance angle defined as
\begin{equation}
\phi = 3\lambda_{SE} -2 \lambda_{J} - \varpi_{SE}.
\end{equation}
Here $\lambda_{SE}$ and $\lambda_{J}$ are the instantaneous mean longitudes
of the Super-Earth and Jupiter respectively, and $\varpi_{SE}$ is the
longitudes of periastron of the Super-Earth.
For given initial values of the eccentricity and resonance angle the motion
of planets will follow the closed curve in this plane.
In Fig.~\ref{fig3} we have shown three different phase space trajectories
for the three different Jupiter locations at 1.27, 1.28 and 1.30. If the
Jupiter is at 1.27, the Super-Earth
has a regular very low eccentricity orbit (first panel), if it is placed
at 1.30 the Super-Earth trajectory librates around the
stable equilibrium point in the 3:2 resonance zone (third panel), and 
in between (when the Jupiter is at 1.28)
the planet experience a chaotic behaviour. 
Similar chaotic zone has been 
found by \cite{holmur} in the asteroid belt. They have pointed out 
the importance of higher order resonances for the origin of chaos.
In this context is worth mentioning the existence of 13:9 commensurability
in the vicinity of this chaotic zone. 
This part of our investigations
has been also summarized in Table~\ref{tab1}.

In the presence of gas giant, the dynamics of the Super-Earth with
a mass of 5.5 $M_{\oplus}$ is similar to that of the massless body 
in the restricted three-body problem. This gives us an opportunity to
determine some of the properties of its
motion by means of analytical, well understood
methods
\citep{murder}.
The extensive discussion of the classical three-body problem (Solar-type
star, Jupiter and Super-Earth) brought us
to the stage where we can predict stability and the character of the planetary
orbits. At the same time we have prepared the well defined framework for
our main study, which aim is to determine how this picture will be
modified in a young planetary system in the
 presence of the gaseous protoplanetary
disc in which the planets are embedded. 


\section{Three-body  problem with a gaseous disc}

\subsection{Description of the performed simulations}
\label {section41}
The system described in the previous Section has been considered again, but
this time the planets are embedded in a gaseous disc. We investigate the
orbital evolution of the Super-Earth and Jupiter in the same region as 
before taking into account the disc-planet interaction. The disc 
has a constant aspect ratio $\frac{H}{r}=0.05$. It extends from 0.33 till
4 (the unit of distance is the initial semi-major axis of the Super-Earth as
it has been chosen in the previous Section). The kinematic viscosity is 
constant in space and its value is 
 $\nu=10^{-5}$ in units of $a_{SE}^2 (GM_*/a_{SE}^3)^{1/2}$, where $a_{SE}$
is the initial semi-major axis of the Super-Earth and $G$ is
the gravitational constant. 
This value, with a disc aspect ratio of 0.05
should correspond to $\alpha =4\cdot 10^{-3}$ at the distance 1 in
our units in the standard 
$\alpha$-parametrization for the viscosity in the disc. 
The planets has been placed in the flat part of the 
surface density
profile with $\Sigma=2 \cdot 10^3 (5.2 {\rm au}/a_{SE})^2$ kg m$^{-2}$:
the standard value attributed to the minimum  mass solar
nebula at 5.2 au  when $a_{SE}=5.2$ au. 
The initial eccentricity of both planets is set to zero. 
We perform numerical simulations using hydrodynamical code NIRVANA,
for details of the numerical scheme and code adopted, see \cite{nelson2000}. 
We have
used the resolution $384\times512$ grid-cells in the radial and the 
azimuthal direction respectively.
We are aware of the fact that adopted resolution might be not high enough 
to simulate properly the evolution of the Super-Earth in general, because 
the corotation region around planet is covered only by a few grid cells. 
However, we think that adopted resolution is a good compromise between the
accuracy of the simulations and reasonable computational time.
The computational domain is a complete ring where the azimuthal 
coordinate
$\varphi =0$ corresponds to $\varphi =2\pi$.
We put the open boundary conditions in  
the radial direction. The  potential is softened with the parameter 
$b=0.04r$ in such a way that the $1/r$ law has been changed into
$1/\sqrt{r^2 + b^2}$.


\begin{table*}
\centering
\begin{minipage}{160mm}
\caption{\label{tab1}
{The initial separation of planets in Super-Earth units (in Jupiter units)
and their final configurations.
The double line denotes the Hill stability
boundary.}}
\begin{tabular}{@{}cccccccccc@{}}
\hline
Initial Super-Earth & Initial Jupiter &Final outcome of the evolution
& Final outcome of the evolution   \\
location & location & without disc & with disc \\
\hline
1.0 (0.870) & 1.15 (1.0) & scattering (before first conjunction) &  scattering  \\
 \hline
1.0 (0.862) & 1.16 (1.0)  & scattering (before first conjunction) &  scattering  \\
 \hline
1.0 (0.855) & 1.17  (1.0) & scattering (before first conjunction) & scattering   \\
 \hline
1.0 (0.847) & 1.18  (1.0) & scattering (after 39 conjunctions)    &  scattering         \\
 \hline
1.0 (0.840) & 1.19  (1.0) & scattering (after 20 conjunctions)    & scattering   \\
 \hline
1.0 (0.833) & 1.20  (1.0) & scattering (after 2 conjunctions)     &  4:3         \\
 \hline
1.0 (0.826) & 1.21  (1.0) & scattering (after 15 conjunctions)    & scattering   \\
 \hline
1.0  (0.820) & 1.22  (1.0) & scattering (after 9 conjunctions)     & 4:3           \\
 \hline
1.0 (0.813) & 1.23  (1.0) & scattering (after 107 conjunctions)   & 4:3           \\
 \hline
1.0 (0.806) & 1.24  (1.0) & scattering (after 5 conjunctions)     & 4:3           \\
 \hline
 \hline
1.0 (0.800) & 1.25  (1.0) & survival (more than 100,000 conjunctions) &   4:3  \\
 \hline
1.0 (0.794) & 1.26  (1.0) & survival (more than 100,000 conjunctions)& 3:2, scattering  \\
 \hline
1.0 (0.787) & 1.27  (1.0) & survival (more than 100,000 conjunctions) & 3:2, scattering \\
 \hline
1.0 (0.781) & 1.28  (1.0) & survival (more than 100,000 conjunctions) & 3:2\\
    &      &         chaotic orbit                       &    \\
 \hline
1.0 (0.769) & 1.30  (1.0) & survival (more than 100,000 conjunctions) &  3:2  \\
 \hline
1.0 (0.741) & 1.35  (1.0) & survival (more than 100,000 conjunctions) &  3:2  \\
 \hline
1.0 (0.690) & 1.45  (1.0) & survival (more than 100,000 conjunctions) &  3:2  \\
\hline
 \end{tabular}
  \end{minipage}
   \end{table*}

As in the case described in previous Section, where the disc was absent
we have investigated $5.5M_{\oplus}$ and $1M_J$ planets with the initial
positions specified in Table~\ref{tab1}. The Super-Earth is located initially 
at
the distance 1 in each simulation. The Jupiter mass planet is located on
the external orbit at the different locations.
Such configuration allows to attain convergent migration
(Fig.~\ref{fig1}).

\subsection{Super-Earth in  mean motion resonance with Jupiter}

In agreement with the theoretical predictions (Fig.~\ref{fig1}) for 
our particular
choice of the system parameters, the Jovian mass planet migrates faster than 
the Super-Earth, and thus planets approach each other and may become eventually
captured in a mean motion resonance.
In order to investigate how an outcome of such evolution depends on the
initial configurations of planets we have performed our simulations 
with a wide range of the planetary separations (see Table~\ref{tab1}). 
We are particularly interested in the conditions for which the
first order mean motion resonances 
$p+1:p$ for $p \ge 2$ are attained (here $p$ is
an integer).
The 2:1 commensurability was not taken into account because 
of the computational time constraints.
The  resonances located closer to the Jupiter like 
5:4, 6:5 and with higher $p$ are not possible,
because for such small separations the system become unstable 
and Super-Earth is
scattered from the disc. 
We have found that two outcomes are
possible, namely the Super-Earth is ejected from the disc or planets 
become locked
in 3:2 or 4:3 mean motion
resonances. 

\begin{figure*}
\begin{minipage}{90mm}
\centering
\vspace{80mm} 
\includegraphics{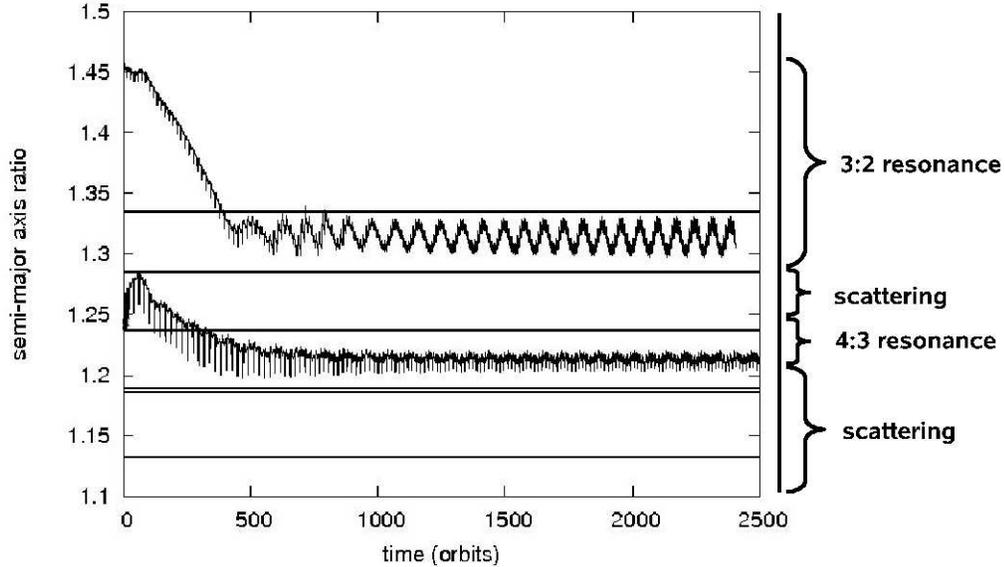}
\caption{\label{fig4}{
The evolution of the ratio of semi-major axes for the Jupiter and Super-Earth
embedded in a gaseous disc. In the case of upper curve 
the initial semi-major axis ratio is 1.45 and planets became locked into
3:2 resonance. For the lower curve, the initial semi-major axis ratio is
1.24 and the attained commensurability is 4:3. The solid horizontal lines
are the 3:2, 4:3 and 5:4 resonance widths in 
the circular three-body
problem. Note that 4:3 and 5:4 resonances partially overlap each other. 
 }}
\end{minipage}
\end{figure*}
\begin{figure}
 \vskip 6.5cm
 \includegraphics{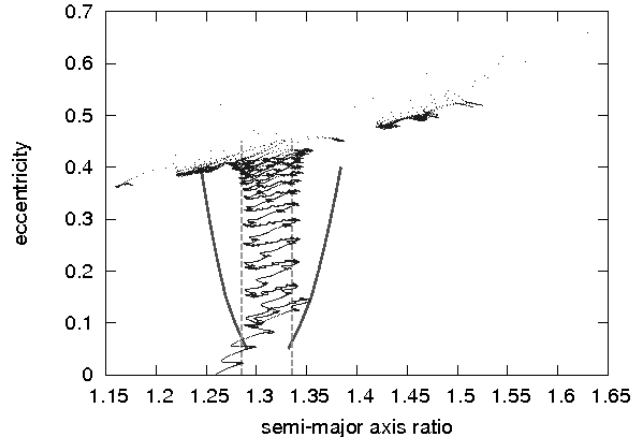}
 \caption{\label{fig5}{The Super-Earth eccentricity excitation for the 
case of Jupiter located at 1.26. The amplitude of the semi-major axis ratio
oscillations around the position of the 3:2 resonance increases
and eventually the Super-Earth is scattered from the disc. The dashed
line is the zero eccentricity resonance width and the solid line
is the libration width obtained in the pendulum approximation for 3:2
resonance (the same as in Fig~\ref{fig2}).}}
  \end{figure}

Our results has been summarized in Table~\ref{tab1} column 4 and on the
right hand side of the Fig.~\ref{fig4} where the most likely outcome of the
planet evolution is given for every initial semi-major axis ratios.
In Fig.~\ref{fig4} we show also two examples of the resonant trapping
occurred for the relative planet separation 0.45 (upper curve) and
0.24 (lower curve). In the first case, the differential migration brought
planets into 3:2 commensurability and in the second one into 4:3.
The semi-major axis ratio
of planets librates exactly within regions of the resonance width marked
by the  solid, horizontal lines in Fig~\ref{fig4}. We plot there the resonance 
widths for $(p+1):p$ commensurabilities with $2\le p \le 4$
in a case of circular orbits 
in the restricted three body problem where the secondary is a Jovian 
mass planet \citep {wisdom80, lecar}. The width of 3:2 and 4:3 resonances
have already been illustrated in Fig.~\ref{fig2} and discussed in the
previous Section.
As we can see the $4:3$ and $5:4$ mean motion resonances
partially overlap each other, which 
 might have a significant effect on  the  stability 
of the system. Indeed  planets located initially in this region 
were immediately scattered from the disc.
Both commensurabilities shown in Fig.~\ref{fig4} are stable for the whole
time of our simulations. We have followed their evolution for not longer than
2500 orbits (the unit of time is defined as the orbital period on the
initial orbit of the Super-Earth divided by 2$\pi$) because after that time 
the surface density is significantly reduced due to viscous evolution
of the disc. However, not all planetary pairs initially locked in the
mean motion resonance survived further evolution (always limited to
2500 orbits). In the case of 3:2 commensurability the libration width
increases with time. The amplitude of the oscillations is bigger for planets
with a separation smaller than the lower boundary of the 3:2 resonance 
region. So in fact in the case of Jupiter placed at 1.26 and 1.27 the 
amplitude grows and eventually the Super-Earth is scattered from the
disc. We have illustrated it for the Jupiter at 1.26 in Fig.~\ref{fig5}.
The direct reason of this scattering is the excitation of the Super-Earth 
eccentricity to the high value, which favours close encounters and ends
with the ejection of the planet. 

\begin{figure}
\vskip 6.5cm
\includegraphics{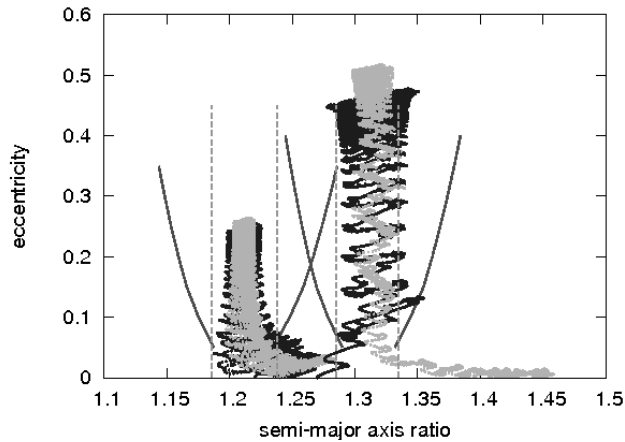}
\caption{\label{fig6}{
Four examples of the Super Earth eccentricity evolution.
For the 4:3 commensurability (left) we are presenting the eccentricity 
as a function
of semi-major axis ratio for the planets with their initial separation
0.24 (gray colour) and 0.22 (black colour) and for the 3:2 commensurability 
(right) the initial planet separation reads 0.45 (gray colour) and 0.27 (black
colour).
The resonance widths are the same as in Fig.~\ref{fig2}.
}}
\end{figure}

We have also investigated eccentricity evolution of the Super-Earth
in few other cases.    
Fig.~\ref{fig6} 
shows eccentricity of the smaller planet versus semi-major axis ratio.
The dashed, vertical lines denote the resonances width for circular case, 
the solid lines
instead
the maximum libration width obtained from the formula given by 
\cite{winmur} in the pendulum approach
in the case of the classical restricted three body problem. We have 
already introduced these quantities in the previous Section.  
The eccentricity of the Super-Earth 
locked in $3:2$ commensurability increases and at the end of simulation 
reaches
value about $0.5$. Eccentricity of planets approaching $3:2$ mean 
motion resonance
from higher value of semi-major axis exceed value of $0.5$. 
For the planets attaining commensurability from the lower value of
the semi-major axis ratio than needed for the exact 3:2 resonance, 
the oscillations around exact
position of the resonance increase and eventually Super-Earth is 
scattered from
the disc (Fig.~\ref{fig5}).

The Super-Earth locked in 4:3 mean motion
resonance in all simulations approach eccentricity value around 0.3. This 
is a lower value of eccentricities than in the case of 3:2 commensurability.
The amplitude libration around the position of the exact resonance is 
smaller than for  3:2 and there is no trend in the amplitude to increase.

\begin{figure}
\vskip 6.5cm
\includegraphics{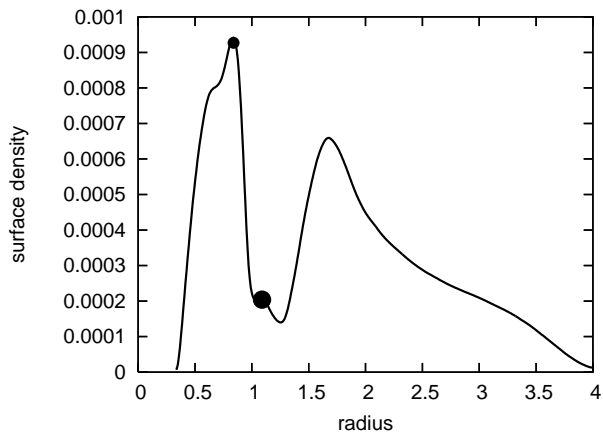}
 \caption{\label{fig7}{The disc profile and planet locations at the
 time of 3:2 resonant capture. The initial position of Jupiter is 1.30.
 }}
 \end{figure}

\begin{figure}
\vskip 6.5cm
\includegraphics{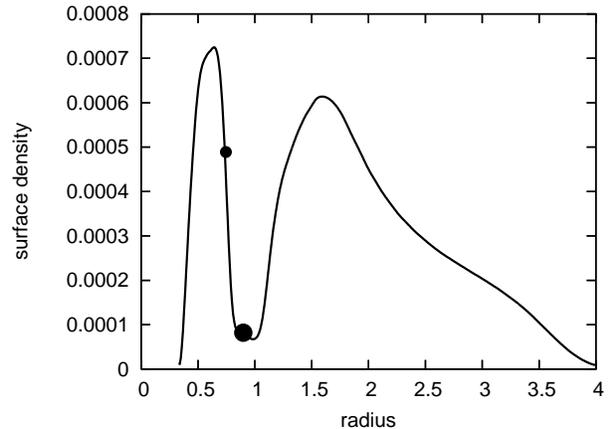}
 \caption{\label{fig8}{The disc profile and planet locations at the time 
 of 4:3 resonant capture. The initial position of Jupiter is 1.24.
 }}
 \end{figure}

Till now we have concentrated on the orbital elements evolution occurring
due to the tidal planet disc interaction. Another effect of this interaction
in the case of the giant planet is the gap opening process, which changes
the surface density profile and may affect the Super-Earth migration.
 In Fig.~\ref{fig7} 
we show the surface density profile together with the positions of 
planets at the moment when they approach $3:2$ mean motion resonance.
In this simulation the Jupiter is placed at 1.30 which is close to the
location of the nominal resonant semi-major axes. After about 628 orbits
the resonant behaviour is clearly
seen in both resonance angles. The gap is still not entirely opened. The
Jupiter is located in it and the Super-Earth is right at the edge of the gap. 
In Fig.~\ref{fig8} we plot the same as in Fig.~\ref{fig7} but for $4:3$
commensurability. Here the Jupiter started at 1.24 and 
the resonant capture happened after about 1178 orbits.
It can be seen that at the time of locking in $4:3$ mean motion resonance 
the Super-Earth orbits in
lower surface density region in the steep gradient of surface density
at the wall of the gap.

The interplay between the migration of both planets and ongoing
process of the gap opening in the disc leads to the resonant
configurations for a range of the initial planet locations. 

\section {Discussion and conclusions}
\label{section5}
 In this paper we have investigated the evolution of Super-Earth and Jupiter
mass planets embedded in gaseous protoplanetary discs where terrestrial
planet is moving on the internal orbit and gas giant on the external one. 
We have found that
according to the simple theoretical prediction
different rates of migration lead to the planet capture in
the mean motion
resonances. For the system considered here the only possible first order
commensurabilities 
are with $p\le3$. For higher $p$ the resonance locations will be in the
Hill unstable zone and in the regions where the resonances begin to overlap
therefore the  
system will be  unstable.

The essential feature of our scenario or differently, 
the reason for the occurrence of
the mean motion resonances in our system, is the convergent migration of 
planets embedded
in the protoplanetary discs. We have mentioned already in 
Section~\ref{section2} that for the given masses and disc profile we can
control the convergence of the migration by the viscosity in the parent
disc, at least for the standard range of parameters, as in the case considered
here. In general, the situation may be more complex. First of all, it has
been found that the torque acting on the low-mass planets might be considerably
smaller than the analytic linear estimate 
and its value does depend on the viscosity
(see \cite{masset} for the discussion). Moreover, it has been also proposed
that even the planets as small as Super-Earths can open a gap in the disc and
slow down their orbital  migration (\cite{matsumura} and references therein). 
Finally, combining the 
equation (\ref{gap}) with the
gap opening formula suggested 
for a disc with a very small viscosity \citep{linpap93},
\cite{crida} constructed the generalized criterion  which is
more restrictive than equation (\ref{gap}). It is than required to
consider in detail the disc profile, masses of the planets and
their locations in the disc in order 
to determine whether or not  we might expect the convergent migration.

Our results for the case of the mature planetary systems when the disc 
is absent are consistent with those predicted analytically for the 
classical planetary  three-body problem. 
From the direct orbit integration we have found the Hill stability region 
which occurs when the Jupiter and Super-Earth orbit relative separation
is 0.24 or bigger, 
which is exactly what we get from the formula (\ref{glad024}).
The chaotic orbits have been identified in the configurations with the
initial semi-major axis ratios in between 1.277 and 1.289. This chaotic
zone might be connected with 13:9 resonance, which partially overlap
with the 3:2 commensurability. 

As has been summarized in Table \ref{tab1} the planets embedded in the 
disc show the same behaviour as planets in matured planetary system (i.e.
where the disc is absent) for the relative separation between planets
less than 0.20. So for these separations the presence of the gas in the
system does not change the outcome of the evolution. The planets are
unstable in the Hill sense \citep{gladman} and also they are in the region
with the resonance overlapping \citep{wisdom80}. The first difference between
young and mature  planetary  systems has been found for the relative 
separation 0.20. The Super-Earth has been trapped by Jupiter in 4:3
mean-motion resonance very quickly preventing in this way the scattering.
This was not the case for 0.21 and the Super-Earth has been ejected from the
disc. For the even higher separations from 0.22 till 0.24 which are also
Hill unstable we have obtained the 4:3 resonance. The last 4:3 commensurability 
we have got for the Super-Earth separated from Jupiter  by 0.25. Next two
configurations with separations 0.26 and 0.27, located inside the
chaotic zone of \cite{wisdom80} passed through unstable 3:2 resonance. All
other investigated by us structures, namely for the separations
0.28-0.45 have been locked in
3:2 commensurability. 

The presence of the disc and its 
interactions with the planets
might be a
key to the attaining resonant configurations like those observed between
Jupiter and groups of asteroids in our system. 
The fate of the Super Earth in the case of the presence of Jupiter mass planet 
on the external orbit is determined
by separation of the planets and their eccentricity evolution.
The resonant structures  with 3:2 and 4:3 commensurabilities
are easy to attain but their stability requires further studies.
Our scenario indicates the possibility that terrestrial  planets 
and giant planets coexist in relatively close proximity. It has an 
interesting 
implication for the astrobiological studies. For example, 
if we  consider a system,
which contains a giant planet in or near its habitable zone, 
taking into account  a simple stability analysis as in Section~\ref{section3}, 
we
might conclude that there is no dynamical room for terrestrial planets
in the habitable zone, close to the Jupiter mass planet. 
However, as it has been shown here, if we allow for the possibility of mean
motion resonance locks among planets, this is no longer necessarily the case.
This hypothesis can be verified by the discoveries of Jovian type planets 
in the habitable zones, similar to the low mass
gas planet  which has been found recently by \cite{fischer}
in 55 Cancri system, 
so the intensive search for the low mass planets which may be present close
to the gas giants is more than justified.
The discovery of the resonant configurations  would provide 
constraints also on the migration processes  themselves. 
If the future observations will find primarily smaller planets locked in 
the interior mean motion resonances and not those locked in the exterior
ones as in \cite{thommes}, or other way round, than we will get important 
information on the
characteristics of the protoplanetary disc in which these planets form. 
Finally, the occurrence of mean motion resonances may increase the chance
of a detection of the terrestrial planets. The best example is provided by
the transit timing measurements, which  may detect additional planets
(which are as small as the Super-Earth considered here) in  the system via 
their gravitational interaction with the transiting Jupiter mass planet,
taking advantage of the resonance induced by migration.

\section*{Acknowledgments}
This work has been partially supported by MNiSW grant N203 026 32/3831
(2007-2010).  
The simulations reported here were performed using the   
Polish National Cluster of Linux Systems (CLUSTERIX). 
We are grateful to John Papaloizou for introducing us to this interesting
subject, Franco Ferrari for his invaluable suggestions about the three body
problem and his continuous support in the development of our computational
techniques and computer facilities and Adam {\L}acny for his helpful
discussion.  Finally, we would like to thank the anonymous referee for 
useful comments and suggestions, which improved the paper.

\appendix
\label{lastpage}
\end{document}